# Effect of spin torque on magnetization switching speed having nonuniform spin distribution

Kazushige Hyodo[1], Chiharu Mitsumata[1], Akimasa Sakuma[1]

[1]Department of Applied Physics, Tohoku University, Sendai 980-8579, Japan

**We study the influence of the spin torque, which depends on the space and time derivative of magnetization, on magnetization reversal time in a ferromagnetic fine particle. The spin torque operates to dissipate the angular momentum of the magnetization precession, and the torque increases in a spin vortex structure. We calculate the magnetization reversal time under a DC magnetic field using the Landau-Lifshitz-Gilbert equation containing a spin torque term. We found that the spin torque changes the magnetization switching speed significantly during the reversal process by maintaining a spin vortex in an intermediate state.**

## I. INTRODUCTION

Detailed clarification of the magnetization dynamics is a key component in spintronics device design.

Understanding the Gilbert damping constant, which characterizes the dissipation of the magnetization precession motion, is particularly significant. Magnetization dynamics are well described by the Landau-Lifshitz-Gilbert (LLG) equation in which the Gilbert damping constant, dependent only on the material, is introduced phenomenologically.

An important factor in the magnetization dynamics of ferromagnetic (FM) metals is the interaction between magnetization and conduction electron spin. In the presence of magnetization precession with an inhomogeneous distribution, the interaction on conduction electron spin can be regarded as the effective electric field and the effective magnetic field, which depends on the conduction electron spin. Here, we focus on this effective electric field ($E_\mu$). When the interaction is sufficiently strong, $E_\mu$ is expressed as follows [1]:

$$E_\mu = \pm \frac{\hbar}{2e}\left(\frac{\partial \boldsymbol{m}}{\partial t} \times \frac{\partial \boldsymbol{m}}{\partial \mu}\right) \cdot \boldsymbol{m} \, , \qquad (1)$$

where $\mu$ represents the Cartesian coordinates x, y and z, the $\pm$ sign corresponds to the $E_\mu$ that the up and down spins receives, $\boldsymbol{m}$ is the magnetization unit vector, $\hbar$ is Plank's constant, and $e$ is the electron charge. $E_\mu$ gives rise to the electric current in the FM metal due to spin polarization. Therefore, the existence of $E_\mu$ is confirmed experimentally by measuring the electromotive force of $E_\mu$ [2]-[4].

In contrast, our study pays attention to the spin current induced from $E_\mu$. Divergence of the spin current generates the spin torque as follows,

$$\boldsymbol{\tau} = -\nabla \cdot \boldsymbol{J}_{\mathrm{s}} \, . \qquad (2)$$

Namely, in FM metals, an inhomogeneous magnetization precession operates on the magnetization dynamics recursively through the conduction electrons. The spin torque works to dissipate the angular momentum of the magnetization precession [5].

The LLG equation including spin torque is expressed as follows [5]:

$$\frac{\partial \boldsymbol{m}}{\partial t} = -\gamma \boldsymbol{m} \times \boldsymbol{H} + \alpha \boldsymbol{m} \times \frac{\partial \boldsymbol{m}}{\partial t} + \eta \sum_{\mu=x,y,z} \frac{\partial \boldsymbol{m}}{\partial \mu}\left[\left(\frac{\partial \boldsymbol{m}}{\partial t} \times \frac{\partial \boldsymbol{m}}{\partial \mu}\right) \cdot \boldsymbol{m}\right] \, , \quad (3)$$

where $\gamma$ is the gyromagnetic ratio, $\boldsymbol{H}_{\mathrm{eff}}$ is the effective magnetic field, and $\eta = (g\mu_{\mathrm{B}}\hbar\sigma)/(4e^2 M_{\mathrm{s}})$. In the expression for $\eta$, $g$ is Landé $g$-factor, $\mu_{\mathrm{B}}$ is the Bohr magneton, $\sigma$ is the electric conductivity, and $M_{\mathrm{s}}$ is the saturation magnetization. The third term on the right-hand side in (3) is the spin torque which contains space and time derivative. In inhomogeneous configurations of the magnetization, the spin torque term will dominate the magnetization damping compared with the Gilbert damping in the second term in (3). For instance, the spin torque alters the frequency of the magnetization rotational motion in current-induced domain walls according to the value of $\eta$ and the width of the domain wall, that is, magnitude of magnetization space derivative [6].

In this paper, we investigate the influence of the spin torque generated in inhomogeneous magnetization precessions on the reversal time of the magnetization switching by an external field. Simulations are carried out for a finite sized FM metal in which we can anticipate the spin torque induced by inhomogeneous magnetization distribution in a considerably large demagnetizing field.

## II. SIMULATION MODEL AND MATERIAL PARAMETERS

We assume a cube (30 nm × 30 nm × 30 nm) of permalloy divided into 16 × 16 × 16 cells. The material parameters are given as follows: $M_{\mathrm{s}}$ = 1.0 (wb/m$^2$), exchange stiffness constant $A$ = 1.0 × 10$^{11}$ (J/m), anisotropy energy constant $K_{\mathrm{u}}$ = 1.0 × 10$^2$ (J/m$^3$), Gilbert damping constant $\alpha$ = 0.008[7], and $\eta$ = 0.47 (nm$^2$) [5]. The easy axis is set in the z-direction, and $\boldsymbol{H}_{\mathrm{eff}}$ includes the exchange, crystal anisotropy, demagnetizing and external magnetic fields. The exchange field is expressed by replacing the spatial derivative by difference as

$$\boldsymbol{H}_{\mathrm{ex}}(i,j,k) = \frac{2A}{M_{\mathrm{s}}}\frac{\mathrm{d}^2 \boldsymbol{m}(i,j,k)}{\mathrm{d}x^2}$$
$$= \frac{2A}{M_{\mathrm{s}}}\frac{\boldsymbol{m}(i+1,j,k)+\boldsymbol{m}(i-1,j,k)}{\Delta x^2} \, , \qquad (4)$$

where $i$, $j$, and $k$ represent the coordination of the cell, and $\Delta x$ is the distance between neighboring calculation points. In deriving (4), we diminish $\boldsymbol{m}(i,j,k)$ in the expansion of $\mathrm{d}^2\boldsymbol{m}(i,j,k)/\mathrm{d}x^2$ because the precession term of the LLG equation is $-\gamma\boldsymbol{m}\times\boldsymbol{H}$.

To evaluate the exchange field at the edge of the cube, we assume a boundary condition that permits us to regard the magnetization of a fictitious cell as that of the neighboring cell

$$\boldsymbol{m}(0,j,k)=\boldsymbol{m}(1,j,k) \text{ or } \boldsymbol{m}(0,j,k)=\boldsymbol{m}(2,j,k) \quad . \quad (5)$$

The exchange field at the edge of the cube with respect to both conditions is given by

$$H_{\mathrm{ex}}(1,j,k)=\frac{2A}{M_{\mathrm{s}}}\frac{\boldsymbol{m}(2,j,k)}{\Delta x^2} \quad \text{if} \quad \boldsymbol{m}(0,j,k)=\boldsymbol{m}(1,j,k)$$

$$H_{\mathrm{ex}}(1,j,k)=\frac{4A}{M_{\mathrm{s}}}\frac{\boldsymbol{m}(2,j,k)}{\Delta x^2} \quad \text{if} \quad \boldsymbol{m}(0,j,k)=\boldsymbol{m}(2,j,k). \quad (6)$$

To ensure accuracy in our calculation, we investigate the difference in the magnetization reversal process between the two boundary conditions. Consequently, we calculate the magnetization reversal of the permalloy cubic particle used in our study while changing the calculation cell division number. We then find that the difference becomes sufficiently small when we divide the magnetization calculation cell over the 16 × 16 × 16 cells. Thus, we take this to be the calculation cell division number in our numerical calculation.

In addition, all these effective field (especially, demagnetizing field) can be used in the case of layered film structure, that enables us to simulate the actual device form.

To investigate the impact of spin torque, the numerical calculation has been carried out using LLG equation: $\eta = 0.47$ (nm$^2$) and $\eta = 0$ in (3); this is equivalent to considering the problem with and without spin torque. The LLG equation is solved using the forward difference method with a time step of $\Delta t = 0.1$ ps. For the initial condition, the magnetization in all calculation cells is defined to be in the z-direction, which is parallel to the easy axis. The magnetization is relaxed under zero field to reach an equilibrium state called the "flower state." Next, at t = 0, a DC magnetic field $\boldsymbol{H}_{\mathrm{a}}$ is applied in the −z-direction. The magnetization reversal process is classified into three patterns depending on the strength of the external DC field. Therefore, our calculation for each $\eta$ is performed for each reversal pattern, in particular $\boldsymbol{H}_{\mathrm{a}} = 0.035/\mu_0$ (A/m), $0.1/\mu_0$ (A/m), and $0.2/\mu_0$ (A/m), where $\mu_0$ is the magnetic permeability in vacuum.

## III. RESULTS AND DISCUSSION

### A. Dependence of Magnetization Reversal Process on External Field

The switching field for the cube is approximately $0.026/\mu_0$ (A/m) and the hysteresis loop (not shown) is square. We calculate the z-component of $\boldsymbol{m}$, $\langle m_z \rangle$, averaged over all cells as a function of the applied time of the external field in the magnetization reversal process.

The magnetization reversal shown in Fig. 1 depends on initial magnetization distributions. However, in the periodic switching of external DC fields having symmetric strength, the magnetization reversal process is reversible without any changes on switching time.

In Fig. 1(a), with $\boldsymbol{H}_{\mathrm{a}} = 0.035/\mu_0$ (A/m), $\langle m_z \rangle$ changes from 1 to -1 between 800 and 900 ns. Once the magnetization reversal starts, $\langle m_z \rangle$ maintains smooth change until the end of the reversal process. There is also little difference in the switching times of the $\eta = 0.47$ (nm$^2$) and $\eta = 0$ cases.

In Fig. 1(b), with $\boldsymbol{H}_{\mathrm{a}} = 0.1/\mu_0$ (A/m), particular emphasis is paid to the magnetization reversal time for the $\eta = 0.47$ (nm$^2$) case as it is reduced to 70% of that in the $\eta = 0$ case. The magnetization reversal occurs in two steps, and the intermediate state $\langle m_z \rangle \cong 0.4$ exists for both value of $\eta$. One

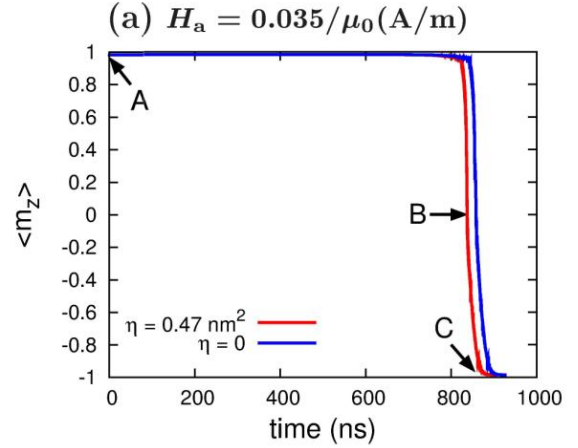

(a) $H_{\mathrm{a}} = 0.035/\mu_0 (\mathrm{A/m})$

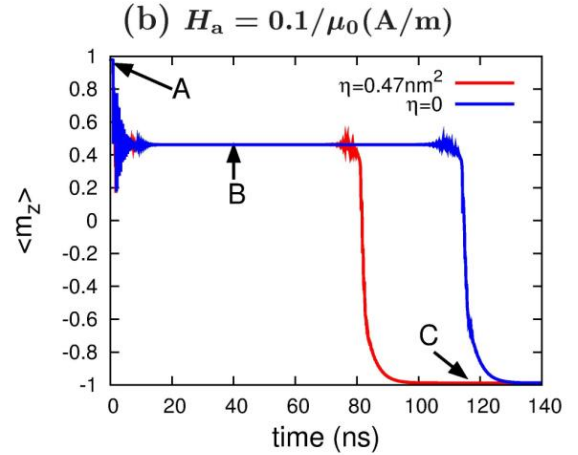

(b) $H_{\mathrm{a}} = 0.1/\mu_0 (\mathrm{A/m})$

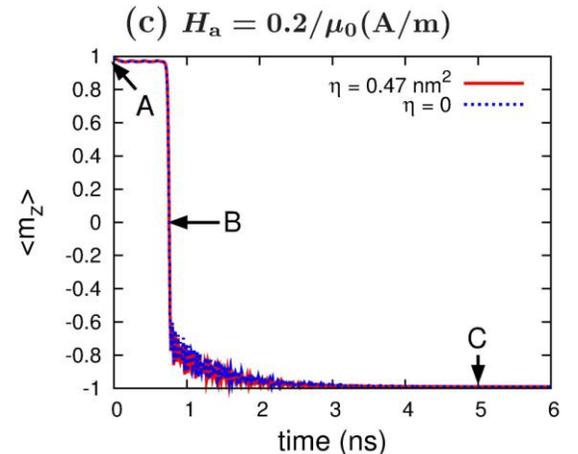

(c) $H_{\mathrm{a}} = 0.2/\mu_0 (\mathrm{A/m})$

Fig.1. [ONLINE COLOR] Plot of the z-component of **m** averaged over all cells $\langle m_z \rangle$, as a function of the applied time of the external field.

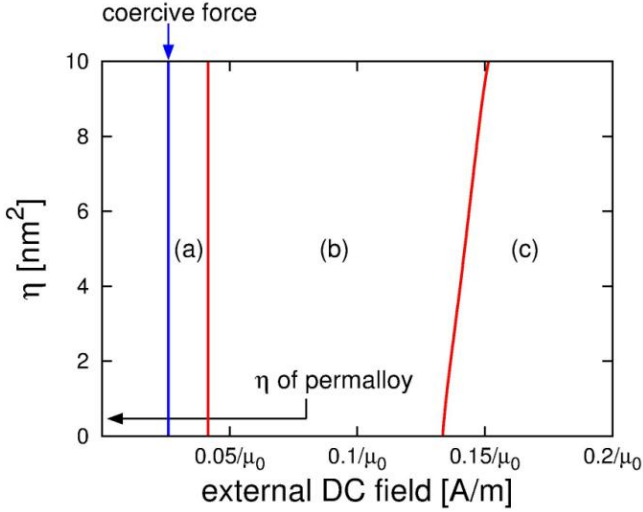

Fig.2. [ONLINE COLOR] A phase diagram of magnetization reversal patterns (a), (b), and (c) shown in Fig.1, which depends on the strength of external DC field and the value of $\eta$ characterizing strength of the spin torque. The transition of each reversal pattern is noncontiguous as drawn by red line. When external field is over $0.2/\mu_0$ (A/m), the magnetization reversal pattern is also (c).

can find that the difference in the breaking time of the intermediate state corresponds to a decrease in the switching time.

In Fig.1 (c), a field of $H_a = 0.2/\mu_0$ (A/m) is applied. The switching time is very short compared with Fig. 1(a) and (b), and $\langle m_z \rangle$ changes rapidly from 1 to $-0.8$ in under 1 ns without entering an intermediate state.

The fluctuation shown in Fig. 1(b) and (c) is caused by small value of Gilbert damping constant $\alpha = 0.008$. Magnetizations seem to fluctuate around stable distribution, particularly after rapid reversal process in Fig. 1(b) and (c). The frequency of fluctuation is collective movement induced by the interaction of many modes of precessions.

Based on the result shown in Fig. 1(a), (b), and (c), we distinguish three of reversal patterns in the phase diagram of Fig. 2, as a function of the external field and $\eta$. The boundary between (a) and (b) is independent of the external field. On the other hand, for the boundary between (b) and (c), the increase of $\eta$ enlarges the critical external field. This change is due to the enhanced magnetization damping by large $\eta$, which works as convergence to the intermediate state.

### B. Effect of Magnetization Distribution on Spin Torque

The strength of external DC field alters reversal process as seen in Fig. 1(a)-(c), and therefore, the spin torque induced from each of these reversal processes is quite different. In this section, we discuss the spin torque effect in terms of the magnetization distribution during the magnetization reversal process. We identify magnetization distributions in three reversal processes, that before reversal, flipping, and after reversal about Fig. 1(a)-(c). We present magnetization distributions only for $\eta = 0.47$ (nm$^2$) in Fig. 3(a)-(c), and when $\eta = 0$, the same distributions appear in the three reversal processes about each magnetization reversal.

The magnetic structure when $H_a = 0.035/\mu_0$ (A/m) is shown in Fig. 3(a) as a cross section image in the x-y plane, which is perpendicular to the external field. (We choose an even $8 \times 8 \times 8$ vector grid in the $16 \times 16 \times 16$ calculation area.) The magnetization flips smoothly with an almost uniform distribution after 800 ns in Fig. 3(a). The spin torque term is quite small compared with the Gilbert damping term in all the reversal process, and consequently, the spin torque scarcely changes the magnetization reversal time.

When $H_a = 0.1/\mu_0$ (A/m), however, $\langle m_z \rangle$ stays constant in an intermediate state until reversal as shown in Fig. 1(b). The magnetic structure of this intermediate state is shown in Fig. 3(b) and shows a spin vortex structure that is almost fixed until the magnetization reversal starts, and the breaking of the intermediate state dominates the magnetization reversal time as shown in Fig. 1(b).

To estimate the strength of spin torque for damping in the spin vortex state in Fig. 3(b), we separate the spin torque into a component parallel to the Gilbert damping term and a component parallel to the precession term. The component parallel to the Gilbert damping term is expressed as follows:

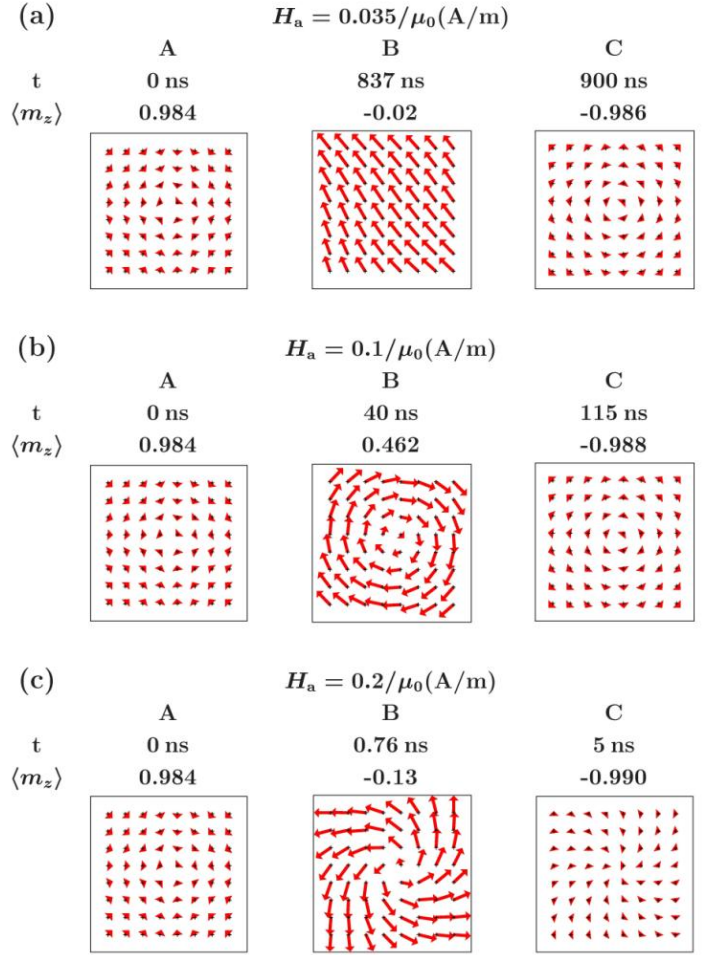

Fig.3. [ONLINE COLOR] (a), (b), and (c) is the cross section image of magnetic structure in x-y plane in Fig.1 (a), (b), and (c). Vectors indicate the x-y component of **m**. Figure A, B, and C are the magnetic structure before reversal, flipping, and after reversal of each magnetization reversal in Fig.1.

t = 40 ns
$m_z = 0.462$

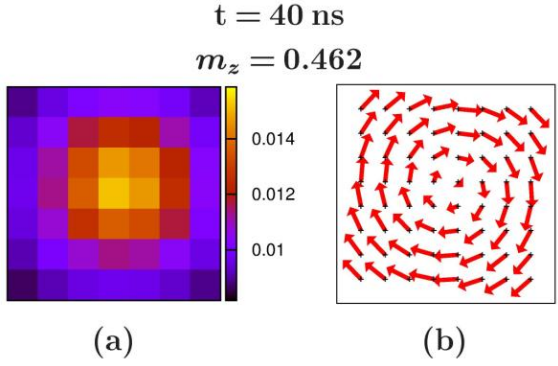

(a)                (b)

Fig.4. [ONLINE COLOR] (a) shows the increased damping constant, which is added to the spin torque component of (8), parallel to the Gilbert damping term whose constant $\alpha = 0.008$. The increased damping constant is calculated from the spin vortex state shown in (b) that appears in the intermediate state in Fig. 1(b).

$$\left( \eta \sum_{\mu=x,y,z} \frac{\partial \mathbf{m}}{\partial \mu} \left[ \left( \frac{\partial \mathbf{m}}{\partial t} \times \frac{\partial \mathbf{m}}{\partial \mu} \right) \cdot \mathbf{m} \right] \right) \cdot \frac{\mathbf{m} \times \frac{\partial \mathbf{m}}{\partial t}}{\left( \frac{\partial \mathbf{m}}{\partial t} \right)^2} = \eta \sum_{\mu=x,y,z} \left( \frac{\partial \mathbf{m}}{\partial \mu} \right)^2 (\sin \theta_\mu)^2 \text{ ,(7)}$$

Where $\theta_\mu$ is the angle between $\partial \mathbf{m} / \partial \mu$ and $\partial \mathbf{m} / \partial t$. With respect to the vortex state as an intermediate state in which magnetization distribution is almost fixed during one precession cycle, the average of (7) over one precession cycle is given by

$$\eta \sum_{\mu=x,y,z} \left( \frac{\partial \mathbf{m}}{\partial \mu} \right)^2 \frac{\int_0^{2\pi} (\sin \theta_\mu)^2}{2\pi} d\theta = \frac{\eta}{2} \sum_{\mu=x,y,z} \left( \frac{\partial \mathbf{m}}{\partial \mu} \right)^2 . \quad (8)$$

The spin torque of (8) corresponds to an addition to α. Thus, we plot the distribution of increased damping constant in Fig. 4(a) with respect to the spin vortex state of Fig. 3(b) (also shown in Fig. 4(b)). The spin torque is comparable to the Gilbert damping constant, especially at the core of vortex due to the high curvature of magnetization in the x-y plane. This increased damping constant, which is larger than $\alpha = 0.008$, breaks the intermediate state earlier and reduces the magnetization reversal time.

When $\boldsymbol{H}_a = 0.2/\mu_0$ (A/m), the vortex state appears again in Fig. 3(c) during the fast magnetization reversal process that occurs in under 1 ns in Fig. 1(c). Although a large spin torque term that is comparable to the Gilbert damping term is expected in this case, the spin torque does not necessarily change the magnetization reversal time. In fact, the magnetization reversal is dominated by the magnetization precession. When a large external field is applied, the effective magnetic field shifts considerably away from the equilibrium position and generates a large magnetization precession radius. This displacement produces energy excitation of the magnetization, and the magnetizations finally reverse. Consequently, the magnetization reversal time does not change, although a large spin torque for magnetization damping is generated.

## IV. SUMMARY

We have demonstrated the influence of the spin torque, generated by inhomogeneous magnetization precession motion in ferromagnetic metals, on the magnetization reversal times of a permalloy cubic particle. The magnetization reversal process can be classified into three patterns depending on the strength of external DC field. When the external field is weak, the spin torque is weak in changing the reversal time due to the uniform magnetization distribution in entire reversal process. When the external field is medium strength, the reversal process has an intermediate state maintaining a spin vortex state. The reversal time is reduced by the spin torque induced from the spin vortex state. The spin torque is as large as the Gilbert damping constant $\alpha = 0.008$ of the permalloy over the long time that the intermediate state is maintained. When the external field is strong, the spin torque does not change the reversal time; however, the vortex state appears in the magnetization reversal. Instead, the magnetization precession term in the LLG equation dominates the magnetization reversal time comparing to the magnetization damping term in the strong external field case.

As a whole, the spin torque does in fact significantly change the magnetization reversal time of magnetization reversal processes with nonuniform distributions.

## REFERENCES

[1] G. E. Volovik, "Linear momentum in ferromagnets," *J. Phys. C: Solid State Phys.*, vol. 20, L83, 1987.

[2] P. N. Hai, S. Ohya, M. Tanaka, S. E. Barnes, and S. Maekawa, "Electromotive force and huge magnetoresistance in magnetic tunnel junctions," *Nature*, vol. 458, pp. 489-492, Mar. 2009.

[3] S. A. Yang, G. S. D. Beach, C. Knutson, D. Xiao, Q. Niu, M. Tsoi, and J.L. Erskine, "Universal Electromotive Force Induced by Domain Wall Motion," *Phys. Rev. Lett.*, vol. 102, 067201, Feb. 2009.

[4] M. V. Costache, M. Sladkov, S. M. Watts, C. H. van der Wal, and B. J. van Wees, "Electrical detection of spin pumping due to precessing magnetization of a single ferromagnet," *Phys. Rev. Lett.*, vol. 97, 216603, Nov. 2006.

[5] S. Zhang, and S. -L. Zhang, "Generalization of the Landau-Lifshitz-Gilbert equation for conducting ferromagnets," *Phys. Rev. Lett.,*. vol. 102, 086601, Feb. 2009.

[6] S-Il. Kim, J-H. Moon, W. Kim, and K-J. Lee, "Current-induced oscillation of a magnetic domain wall: Effect of damping enhanced by magnetization dynamics," *Current. Appl. Phys.*, vol. 11, pp. 61-64, Jan. 2011.

[7] J. P. Nibarger, R. Lopusnik, and T. J. Silva, "Damping as a function of pulsed field amplitude and bias field in thin film Permalloy," *Appl. Phys. Lett.*, vol. 82, 2112, Mar. 2003.